\documentstyle[12pt]{article}
\makeatletter
\@addtoreset{equation}{section}
\makeatother

\newcommand{\bibi}{\bibitem}

\def\a{\alpha}
\def\b{\beta}
\def\c{\chi}

\def\e{\epsilon}   
\def\g{\gamma}
\def\h{\eta}

\def\j{\psi}
\def\k{\kappa}     

\def\m{\mu}
\def\hm{\hat{\mu}}
\def\n{\nu}
\def\o{\omega}

\def\J{\Psi}

\def\O{\Omega}

\def\Jb{\overline{\J}}
\def\jb{\overline{\j}}
\def\cb{\overline{\c}}

\newcommand{\del}{\partial}
\newcommand{\av}[1]{\mbox{$\langle #1 \rangle$}}

\newcommand{\half}{\mbox{{\normalsize $\frac{1}{2}$}} }
\newcommand{\quart}{\mbox{{\small $\frac{1}{4}$}} }

\newcommand{\psnrm}{\mbox{{\normalsize $\frac{1}{2\sqrt{2}}$}} }

\newcommand{\RE}{\mbox{Re\,}}
\newcommand{\IM}{\mbox{Im\,}}
\newcommand{\Tr}{\mbox{Tr\,}}
\newcommand{\ra}{\rightarrow}

\newcommand{\dg}{\dagger}

\newcommand{\al}{\alpha}

\newcommand{\lag}{\langle}
\newcommand{\rag}{\rangle}
\newcommand{\gm}{\gamma}

\newcommand{\dl}{\delta}
\newcommand{\ep}{\varepsilon}

\newcommand{\zt}{\zeta}
\newcommand{\et}{\eta}

\newcommand{\kp}{\kappa}

\newcommand{\ta}{\tau}

\newcommand{\ch}{\chi}

\newcommand{\om}{\omega}

\newcommand{\Ps}{\Psi}

\newcommand{\Om}{\Omega}
\newcommand{\Psb}{\overline{\Ps}}

\newcommand{\Dsl}{D\!\!\!\!/}

\newcommand{\dmu}{\partial_{\mu}}

\newcommand{\mh}{\hat{\mu}}

\newcommand{\hmu}{\hat{\mu}}

\newcommand{\aleq}{\mbox{}^{\textstyle <}_{\textstyle\sim}}

\newcommand{\be}{\begin{equation}}
\newcommand{\ee}{\end{equation}}
\newcommand{\bea}{\begin{eqnarray}}
\newcommand{\eea}{\end{eqnarray}}
\newcommand{\eq}{\ref}
\newcommand{\beq}{\begin{equation}}
\newcommand{\eeq}{\end{equation}}
\newcommand{\cc}{\cite}
\newcommand{\lb}{\label}

\def \3{\ss}

\def\dateandnumber(#1)#2#3#4{
\vbox to 18mm{%
     \hbox to \textwidth{ \hspace*{14mm} \hsize=40mm%
            \vbox{%
                 \hbox to 40mm{\large #1 \hss}%
                 \hbox to 40mm{    \hss}%
                 \hbox to 40mm{    \hss}%
                 }%
                 \hss \hsize=80mm%
            \vbox{%
                 \hbox to 80mm{\hss \large #2}
                 \hbox to 80mm{\hss \large #3}
                 \hbox to 80mm{\hss \large #4}
                 }%
            \hspace*{14mm} }%
      \vss
    }
}
\def\titleofpreprint#1#2#3#4{{\LARGE \bf
\vbox to 43mm{%
     \vss
     \hbox to \textwidth{ \hspace*{14mm} \hsize=130mm%
            \hss \vbox{
                      \hbox to 130mm{\hss \LARGE \bf #1\hss}%
                      \hbox to 130mm{\hss \LARGE \bf #2\hss}%
                      \hbox to 130mm{\hss \LARGE \bf #3\hss}%
                      \hbox to 130mm{\hss \LARGE \bf #4\hss}%
                 }%
            \hss \hspace*{14mm} }%
      \vss
    }}
}
\def\listofauthors#1#2#3{{\large
\vbox to 22mm{%
     \vss
     \hbox to \textwidth{ \hspace*{14mm} \hsize=130mm%
            \hss \vbox{
                      \hbox to 130mm{\hss \large #1\hss}%
                      \hbox to 130mm{\hss \large #2\hss}%
                      \hbox to 130mm{\hss \large #3\hss}%
                 }%
            \hss \hspace*{14mm} }%
      \vss
    }}
}
\def\listofaddresses#1#2#3#4#5{{\small
\vbox to 18mm{%
     \vss
     \hbox to \textwidth{ \hspace*{14mm} \hsize=130mm%
            \hss \vbox{
                      \hbox to 130mm{\hss \small #1\hss}%
                      \hbox to 130mm{\hss \small #2\hss}%
                      \hbox to 130mm{\hss \small #3\hss}%
                      \hbox to 130mm{\hss \small #4\hss}%
                      \hbox to 130mm{\hss \small #5\hss}%
                 }%
            \hss \hspace*{14mm} }%
      \vss
    }}
}
\def\abstractofpreprint#1{{\normalsize
  \vbox to 110mm{%
     \vss
     \hbox to \textwidth{\hss \normalsize \bf Abstract \hss}%
     \vspace*{1cm} \normalsize
     #1
     \vss
    }}
}

\def\footnoteitem(#1)#2{
\begin{list}{#1}{\labelwidth4.0mm \leftmargin7.0mm
\labelsep2.5mm \rightmargin7.0mm \parsep0.5ex plus0.2ex minus0.1ex
\itemsep0ex plus0.2ex }
\item #2
\end{list}
}
\begin{document}
\dateandnumber(June 1993)%
{Amsterdam ITFA 93-13}%
{UCSD/PTH 93-13}%
{              }%
\titleofpreprint%
{     Staggered fermions for chiral gauge                     }%
{     theories: Test on a two-dimensional                     }%
{     axial-vector model                                      }%
{                                                              }%
\listofauthors%
{Wolfgang Bock$^{1,\#}$,
Jan Smit $^{1,\&}$ and Jeroen C. Vink $^{2,*}$                 }%
{                                                              }%
{                                                              }%
\listofaddresses%
{\em $^1$Institute of Theoretical Physics, University of Amsterdam,}%
{\em \\ Valckenierstraat 65, 1018 XE Amsterdam, The Netherlands }%
{\em $^2$University of California, San Diego, Department of Physics,}%
{\em 9500 Gilman Drive 0319, La Jolla, CA 92093-0319, USA}%
{\em            }%
\abstractofpreprint{
As a first step towards constructing chiral models on the lattice
with staggered fermions, we study
a U(1) model with axial-vector coupling to an external
gauge field in two dimensions.
In our approach gauge invariance is broken, but it is
restored in the classical continuum limit.
We find that the continuum divergence relations for the vector and
axial-vector currents are reproduced, up to contact terms,
which we determine analytically.
The current divergence relations are also studied numerically for
smooth external gauge fields with topological charge zero.
We furthermore investigate the effect of fluctuating gauge
transformations and of gauge configurations
with non-trivial topological charge.
}
\noindent $\#$ e-mail: bock@phys.uva.nl   \\
\noindent $\&$ e-mail: jsmit@phys.uva.nl \\
\noindent $*$  e-mail: vink@yukawa.ucsd.edu \\
%
%
%
%
\section{Introduction}
A consistent non-perturbative formulation of
a chiral gauge theory is much desired.
Within perturbation theory the quantization of chiral gauge
theories has been studied and it is claimed to be solved
satisfactorily \cc{Be}, although a gauge invariant regulator does not
appear to exist. Various proposals have been made
for a non-perturbative formulation of chiral gauge theories
on the lattice. A problem is the doubling phenomenon: a gauge
invariant model on a regular lattice is necessarily non-chiral in
the sense that each fermion is accompanied by extra
degrees of freedom, the so-called species doublers, which
couple with opposite chiral charge to the gauge fields and render
the theory vector-like. Most of the currently existing lattice
proposals try to eliminate the unwanted species doublers either
by making them very heavy or by tuning their interactions to
zero (for an overview see
ref.~\cc{REV}). One can also use the doublers
as physical degrees of freedom, with
the staggered fermion method [3-5], which is the
strategy followed in this paper. \\
As in perturbation theory in the continuum, the regulated lattice
model violates gauge invariance, but it is restored in the
classical continuum limit. The question is how to restore it in
the quantum theory. One can mimic the continuum methods closely,
by attempting non-perturbative gauge fixing and adding
counterterms to restore gauge invariance \cc{Rome89}. An
alternative approach focuses on a possible dynamical restoration
of gauge invariance \cc{Sm88,SmROME}. The question of dynamical
gauge symmetry restoration will be investigated in an other
publication \cc{BoSm93}. In this paper we shall test the
staggered method in a two-dimensional axial-vector model, where
the staggered fermion fields are coupled to external gauge
fields. A preliminary account of this work has already been
presented in ref.~\cc{BoSm92}.
\section{The target model}
Our continuum target model is
given by the following euclidean action in two dimensions
\be
S=-\int d^2x \left\{
 \jb \g_{\mu} (\partial_{\mu} +i\g_5 A_{\m})\j + m \jb \j \right\}
  \;, \label{CONT}
\ee
where $A_{\mu}$ is an external gauge field and $\g_5=-i \g_1
\g_2$. For the purpose of numerical simulations and also for the
use in tests we have added a bare mass term for the fermions with
mass parameter $m$.
In two dimensions the axial-vector model (\eq{CONT}) can be
rewritten in  vector form with a Majorana mass term, e.g. by a charge
conjugation transformation on the right handed fermion fields. So
it is not really a chiral gauge theory, and for $m=0$ it is
equivalent to the Schwinger model.
However, the technical aspects
of `$\gm_5$' in its lattice version are very similar to truly
chiral gauge theories. The reason for choosing this target model
and not e.~g. a left-handed model, is that the staggered fermion
version of the axial model has a larger lattice symmetry group.

For $m=0$ the action
(\eq{CONT}) is invariant under the local gauge transformations
$A_{\mu}(x)\ra A_{\mu}(x)+i\O(x)\dmu \O^*(x)$,
$\j(x) \ra (\O(x) P_L+\O^*(x) P_R) \j(x)$, $\jb(x) \ra \jb(x) (\O(x)
P_L+\O^*(x) P_R)$, with  $P_{L,R}=(1\mp \g_5)/2$ and $\O(x) \in$
U(1). The action is furthermore
invariant under the global vector symmetry $\j(x) \ra \Om\j(x)$,
$\jb(x) \ra \jb(x) \O^*$.
The vector and axial-vector currents,
$J^V_{\mu} = i\jb \g_{\mu}\j$ and
$J^A_{\mu}= i \jb \g_{\mu} \g_5 \j$, satisfy the classical
divergence equations
\bea
& & \dmu J^V_{\mu}=0,               \label{MO1} \\
& & \dmu J^A_{\mu}=2 m J^P, \label{MO2}
\eea
where $J^P=i \jb \g_5 \j$ is the pseudoscalar density.
%
%
\begin{figure}
\vspace*{7cm}
\caption{ \noindent {\em
Feynman diagrams for
$T_{\m \n}^{VA}(p)$, $T_{\m}^{VP}(p)$,
$T_{\m \n}^{AA}(p)$, and $T_{\m}^{AP}(p)$.
}}
\label{FIG1}
\end{figure}

As is well known these classical relations (\eq{MO1}), (\eq{MO2}) may
be invalidated by quantum effects.
Requiring gauge invariance for $m=0$, eq.~(\ref{MO2}) has to
remain valid but (\ref{MO1}) becomes anomalous and takes the
form
\be
\dmu J^V_{\mu}= 2 i q \;,\;\;\;\; q=\frac{1}{2\pi} F_{12}\;, \label{DIV}
\ee
where $F_{12}(x)=\partial_1 A_2(x) - \partial_2 A_1(x)$ is the field strength
and $q(x)$ is the topological charge density. The topological
charge $Q$ is defined by $Q=\int d^2 x \; q(x)$. Our aim is to
study the divergence equations (\ref{MO1}) and (\ref{DIV}) in the
lattice version of the model.

Let us briefly review the anomaly structure of the current
divergences in the quantum theory.
We evaluate $\langle \dmu J^{V,A}_{\mu}\rangle$ and $\langle
J^{P}\rangle$ in perturbation theory, by expansion in $A_{\mu}$,
which leads to a series of diagrams in which
external $A_{\m}$-lines are attached to a fermion loop.
We concentrate on the diagrams shown in fig.~\ref{FIG1}a, since
diagrams with one external line vanish and
diagrams with more than two external lines are convergent by
power counting. We
shall evaluate these diagrams using a spherical cut-off in
momentum space, because this regularization has analogies to our
staggered fermion formulation introduced later. Such a cut-off violates
gauge invariance, but the desired result is easily obtained by
adding suitable contact terms. For a discussion (including e.g.
the gauge invariant point splitting method, which gives rise to
the additional diagrams in fig.~\ref{FIG1}b) see
ref.~\cc{GrJa72}.

Although the diagrams in fig.~\ref{FIG1}a seem to be
logarithmically divergent, the momentum cut-off gives a finite
answer when it is removed. Fig.~1a leads to
\bea
T_{\mu \nu}^{VA} (p)&\equiv & \int d^4 x \exp(- ixp)
\lag J^V_{\mu}(x) J^A_{\nu} (0) \rag_{A=0} \nonumber \\
&=& \frac{i}{2\pi} \int_0^1 dz \frac{
m^2 \e_{\mu \nu}-z (1-z) (\e_{\mu \a} p_{\a}p_{\nu}
+\e_{\nu \a} p_{\a}p_{\mu})}{
m^2 + z(1-z) p^2   } + C^{VA}_{\m \n}\;,  \label{TVA1}
\eea
where $C^{VA}_{\m\n}$ is a contact term to be determined shortly.%
\hspace{0.1cm}From this expression we get the two Ward identities
\bea
& & ip_{\mu} T_{\mu \nu}^{VA}(p)=\frac{1}{2\pi} \e_{\nu \a} p_{\a}
+ ip_{\mu} C^{VA}_{\m \n}\;,\; \nonumber \\
& & -i p_{\nu} T_{\mu \nu}^{VA}(p)= 2m T_{\mu}^{VP}(p)
-\frac{1}{2\pi} \e_{\mu \a} p_{\a}
-ip_{\nu} C^{VA}_{\m \n} \;, \label{TVA}
\eea
where the amplitude
\be
T_{\mu}^{VP}(p)= \frac{ 1 }{2\pi}
\int_0^1 dz \frac{m \e_{\mu \a} p_{\a}}
{m^2+z(1-z)p^2} \;,\label{TMU}
\ee
results from the evaluation
of the the $J^V_{\mu}$-$J^P$ correlation function at $A_{\mu}=0$
(see fig.~\ref{FIG1}a). These relations show anomalous terms
in both the vector and axial-vector Ward identities. By choosing
the contact terms $C^{VA}_{\m \n}= \pm (i/2\pi) \e_{\mu \nu}$
one may shift the anomaly either to the vector ($+$) or to the
axial-vector Ward identity ($-$).  Since we insist on axial gauge
invariance the contact term is determined as
\be
C^{VA}_{\m \n} = +\frac{i}{2\pi}  \e_{\m \n} \lb{C1},
\ee
giving
\be
ip_{\mu} T_{\mu \nu}^{VA}(p)
     = \frac{1}{\pi} \e_{\nu \a} p_{\a} \;,\;\;\;
       -ip_{\nu} T_{\mu \nu}^{VA}(p)
     = 2m T_{\nu}^{VP}(p) \;.
\ee
For $\langle J^A_{\mu} \rangle$ fig.~1a gives
\bea
T_{\mu \nu}^{AA} (p) &\equiv & \int d^4 x \exp(- ixp)
\lag J^A_{\mu}(x) J^A_{\nu} (0) \rag_{A=0} \nonumber \\
&=&- \frac{1}{\pi} \int_0^1 dz \frac{
(\dl_{\mu\nu} p^2 -p_{\mu}p_{\nu}) z(1-z) + m^2\dl_{\mu\nu} }{
m^2 + z(1-z) p^2} +\frac{1}{2\pi}\dl_{\mu\nu} + C^{AA}_{\mu \nu}\;.
\lb{TAA}
\eea
With the choice
\be
C^{AA}_{\mu \nu} = -\frac{1}{2\pi}\dl_{\mu\nu} \lb{C2}
\ee
we get from this relation the desired Ward identity
\be
ip_{\n}  T_{\mu \nu}^{AA} (p) = 2m T_{\m}^{AP}(p) \;,\;\;\;
T_{\m}^{AP}(p) = -\frac{i}{2\pi}\int_0^1 dz
\frac{mp_{\m}}{m^2 + z(1-z) p^2}, \lb{AAA}
\ee
with $T_{\m}^{AP}(p)$ the $J_{\m}^A$-$J^P$ correlation function
at $A_{\mu}=0$. From the above results for the current correlation
functions one can derive
the divergence equations (\ref{MO2}) and (\ref{DIV}).

We shall show in the following that the symmetry
properties of the model, as they manifest themselves in the
current divergence relations (\eq{MO2}) and (\eq{DIV}), can be
recovered in a staggered fermion version of the model on the lattice.
\section{The staggered fermion model}
We generalize our target model to two flavors, which makes it
somewhat simpler to describe in the staggered fermion formalism.
We introduce $2\times 2$ matrix fermion fields $\J^{\a\k}_x$ and
$\Jb^{\k\a}_x$ on a two dimensional square lattice, where $\a$
and $\k$ are Dirac and flavor indices, respectively. Using these
matrix fields we find after the naive lattice transcription of
the two flavor version of the target model in eq.~(\eq{CONT})
the following action,
\bea
 S \!\!\!&=&\!\!\! -\sum_{x\m} \half\Tr \left\{
      \Jb_x \g_{\m}(U_{\m x}^L P_L + U^R_{\m x}P_R)\J_{x+\hm}
   -  \Jb_{x+\hm} \g_{\m}(U^{L*}_{\m x}P_L + U^{R*}_{\m x}P_R)\J_{x} \right\}
\nonumber\\
     &&\mbox{} - m \sum_x \Tr \left\{ \Jb_x \J_x \right\} \;,
                   \label{MODPSI}\\
  U^L_{\mu x} \!\!\!&=&\!\!\! e^{-iaA_{\mu x}},\;\;\; U^R_{\mu x} = e^{iaA_{\mu
x}},
                                   \label{AX}
\eea
with $a$ the lattice
spacing. We shall use lattice units, $a=1$. The action above
would be gauge invariant if $\J^{\a\k}_x$ and $\Jb^{\k\a}_x$
would be independent degrees of freedom, and this would lead to
fermion doublers with opposite chiralities.

The fermion doublers are situated at the boundary of the
Brillouin zone in momentum space, i.e. at momenta $p_{\mu} =\pm
\pi$. Consider restricting the momenta of the matrix fields such
that $-\pi/2 < p_{\mu} \leq +\pi/2$. Then we loose the fermion
doublers of the matrix fields, but also gauge invariance. It is
clear, however, that in the classical continuum limit, where the
field momenta go to zero (in lattice units),
eq.~(\eq{MODPSI}) goes over into the eq.~(\eq{CONT}) with two
flavors. Hence gauge invariance gets restored in this limit. One
might think that the cut-off in momentum space has to result in a
non-local action. However, it is possible to express the
action in a form that is local, using staggered fermions.

Staggered fermion fields on the lattice, denoted by the
one-component fields $\c_x$ and $\cb_x$, do not carry explicit
flavor and Dirac labels. These labels are
supplied through the doubling phenomenon.
In the classical continuum limit
one recovers the usual Dirac and flavor structure. We make the
connection with the
$\J^{\a\k}_x$ and $\Jb^{\k\a}_x$ by writing \cc{Sm88,SmROME}
\be
\J_x  = \psnrm\sum_b \g^{x+b}\c_{x+b} \;,\;\;\;\;
\Jb_x = \psnrm\sum_b (\g^{x+b})^{\dg}\cb_{x+b} \;,
                        \label{CHIPSI}
\ee
where $\g^{x} \equiv \g_1^{x_1}\g_2^{x_2}$ and the sum runs over the
corners of an elementary lattice square, $b_{\m}=0,1$.
In momentum space we have the relation \cc{Sm92}
\be
\J_{\al\kp} (p) = z(p) \sum_b T_{\al\kp, b} (p)\c(p+\pi b),\;\;\;
-\pi/2 < p \leq \pi/2 \;,
\ee
where $T_{\al\kp,b} = \sum_c \exp(ibc\pi)\g^c_{\al\kp}/4$
is a unitary matrix and $z(p)$ is a
non-vanishing function in the restricted momentum interval.
This shows clearly that in this restricted interval the Fourier components
of the matrix fields are independent.

Having expressed the matrix fermion fields in terms of the
independent $\c_x$ and $\cb_x$ fields, substitution into the action
(\ref{MODPSI}) leads to a {\em local} action.
The indices $\a$ and $\k$ on
$\J$ and $\Jb$ act like Dirac and flavor indices and one
can construct staggered fermion models involving arbitrary
spin-flavor couplings to other fields in a straightforward manner
such that the target models are recovered in the classical
continuum limit. For the Standard Model and Grand Unified
Theories like SO(10) and SU(5) this can be done such that the
staggered fermion symmetry group is preserved \cc{SmROME}.
This invariance
is important for reducing the number of counterterms needed to get
a satisfactory continuum limit \cc{GoSm}.
This strategy of coupling the staggered fermion spin-flavors
has recently been successfully applied to a fermion-Higgs model \cc{UP}.

By working out the trace in (\ref{MODPSI})
one obtains the action in terms of the staggered fermion fields,
\bea
 S  \!\!\!&=&\!\!\! - \frac{1}{2} \sum_{x\m}
      \left\{ c_{\m x} \quart \sum_b \h_{\m x+b}
         (\cb_{x+b} \c_{x+b+\hm} - \cb_{x+b+\hm} \c_{x+b})
\right. \nonumber\\
\!\!\!&&\!\!\!      \left. - s_{\m x} \quart \sum_{b+c=n}\h_{12 x+c}
(\h_{\m x+c}\cb_{x+b} \c_{x+c+\hm}
- \h_{\m x+b}\cb_{x+b+\hmu} \c_{x+c}) \right\}
- m\sum_x \cb_x \c_x \;, \label{MODCHI}
\eea
with the abbreviations $c_{\m x}=\RE U_{\m x}$, $s_{\m x}=\IM U_{\m x}$
and $n=(1,1)$. The sign factors $\h_{1 x}=1$
and $\h_{2 x}=(-1)^{x_1}$ represent the Dirac matrices
$\g_1$ and $\g_2$ and  $\h_{12 x}= \h_{2 x}\h_{1 x+\hat{2}}
= (-1)^{x_1}$ the $i\g_5=\g_1\g_2$. In the classical continuum
limit this action describes two flavors of axially coupled Dirac
fermions.

For a one flavor staggered fermion model we would have used
only one Grassmann variable per site (so-called `reduced' or `real'
staggered fermions). One then defines the
$\cb_x$ fields usually on the even,
and the $\c_x$ fields on the odd lattice sites.
This would require a one link mass term instead of
the simple one site mass term in (\ref{MODCHI}). The
continuum interpretation is in this case somewhat more involved
\cc{Sm92,DoSm}.

The couplings in the $s_{\mu x}$ term in the action (\ref{MODCHI})
are not confined within a
plaquette. For example, for $c=0$ and $b=n$ we have three link
couplings. The action is of course not unique.
According to standard staggered fermion properties,
the three link couplings may be replaced by one link couplings by
shifting the $\c$ or $\cb$ field over two
lattice spacings in the same direction (even shifts).
Such actions are equivalent in the sense that they lead to the same
classical continuum limit, and in the quantum theory
they are
expected to be in the same universality class. It is instructive
to give here a particularly simple alternative to (\eq{MODCHI}),
\bea
S\!\!\!&=&\!\!\!-\sum_{x\mu} \left\{ \bar{c}_{\mu x} \et_{\mu x} \half
(\cb_x\c_{x+\hm} - \cb_{x+\hm}\c_x) \right. \nonumber\\
&&\left. -\bar{s}_{\mu x}\ep_{\mu\nu} \et_{\nu x} \half
(\cb_x\c_{x+\hat{\nu}} + \cb_{x+\hat{\nu}}\c_x) \right\}
-\sum_x m\cb_x\c_x\;,\label{CAN}\\
\bar{c}_{\mu x} \!\!\!&=&\!\!\! \frac{1}{4} \sum_b c_{\mu x-b}\;,\;\;\;
\bar{s}_{\mu x}=\frac{1}{4} \sum_b s_{\mu x-b}\;,
\eea
obtained by using $c=n-b$, the
identities $\et_{12 x+n} \et_{\mu x+n} = \ep_{\mu\nu} \et_{\nu x}$,
$n+\hm= \hat{\nu} + 2\hm$, and the equivalence of e.g.
$\cb_x\c_{x+n+\hat{\nu} + 2\hm-2b}$ with $\cb_x\c_{x+n+\hat{\nu}}$
which differ by even shifts.
A further reduction could be achieved by the replacements
$\bar{c}_{\mu x} \ra 1$ and $\bar{s}_{\mu x} \ra s_{\mu}$
in the above expression,
as the resulting model still has the same classical continuum limit.

A model with all the couplings confined
within a plaquette may be called a
canonical model, as it allows for a canonical construction of the
transfer operator \cc{DoSm,Sm91}. In the following we shall
use, however, the original version as written in
eq.~(\ref{MODCHI}).

For $m=0$ the action (\eq{MODPSI}) appears to be invariant
under the local gauge transformations,
\bea
& & U_{\m x} \ra \O_x U_{\m x} \O^*_{x+\hm}\;, \label{UUU} \\
& & \J_x \ra (\O_x P_L + \O^*_x P_R) \J_x \;,\;\;\;\;
\Jb_x \ra \Jb_x (\O_x P_L + \O^*_x P_R)\;.   \label{PPP}
\eea
However, this gauge invariance is broken because of the momentum space
cut-off on the matrix fields. The four components
of the $\J_x$ and $\Jb_x$ matrix fields are not independent, as is
evident from their expression in terms of the independent
$\c_x$ and $\cb_x$ fields. Therefore we cannot translate the gauge
transformations on $\J$ and $\Jb$ to local transformations on $\c$
and $\cb$, and the action (\ref{MODCHI}) lacks gauge invariance.

The global vector U(1) transformation
\be
\c_x \ra \exp(i\om) \c_x\;,\;\;\; \cb_x \ra \cb_x \exp(-i\om)
\lb{EU1}
\ee
is an exact symmetry of the actions
(\ref{MODCHI}) and (\eq{CAN}). For $m=0$ there is
furthermore a second exact global U(1) invariance, the
`U(1)$_{\ep}$' invariance $\c_x \ra \exp(i\om\ep_x)\c_x$,
$\cb_x \ra \cb_x \exp(i\om\ep_x)$, $\ep_x = (-1)^{x_1 + x_2}$,
which corresponds to a flavor
non-singlet chiral transformation $\J_x\ra \cos\om\, \J_x +
i \sin\om\, \g_5\J_x \g_5$, $\Jb_x\ra \cos\om\, \Jb_x + i\sin\om\,
\g_5\Jb_x \g_5$. These U(1)$\times$U(1) transformations are part
of the U(2)$\times$U(2) global invariance of the classical two
flavor continuum action. In a one flavor staggered fermion
model U(1)$_{\ep}$ would be the only global U(1) symmetry.

We are not concerned here with the full aspects of flavor
symmetry restoration, but concentrate on the flavor singlet
vector and axial-vector currents, $J_{\mu x}^V$ and $J_{\mu x}^A$.
The latter is the gauge current
which should be exactly conserved. The former is the U(1) vector
current which should have the anomaly discussed in the previous
section.
\section{Divergence equations on the lattice}
In this section we determine the contact terms in the
divergence equations for the gauge current $J^A_{\mu}$ and a
flavor singlet U(1) current $J^V_{\mu}$ in our lattice model
(\ref{MODCHI}).

To introduce these currents we
generalize (\eq{MODPSI}), (\eq{AX}) to
include also an external vector field $V_{\mu x}$, writing
\be
U_{\mu x}^L = e^{-iA_{\mu x} -i V_{\mu x}} \;,\;\;\;
U_{\mu x}^R = e^{+iA_{\mu x} -i V_{\mu x}} \;. \label{MODAV}
\ee
Since gauge invariance is broken we expect to have to add
counterterms to the action to restore it in the continuum
limit. From the sect.~2 we expect these to have the form
\be
S_{ct} = \sum_x \left\{ \half C^{AA}_{\mu\nu} A_{\mu x} A_{\nu x}
+ C^{VA}_{\mu\nu}V_{\mu x} A_{\nu x} +
\half C^{VV}_{\mu\nu} V_{\mu x} V_{\nu x} \right\} \;. \label{CT}
\ee
The coefficients $C^{AA}$ and $C^{VA}$ have to be determined such
that in the scaling region the effective action obtained by
integrating out the fermion fields is invariant for $m=0$ under
the gauge transformations (\ref{UUU}) on $A_{\mu x}$. We have
included a $C^{VV}$ term which may be determined such that under
gauge transformations on $V_{\mu x}$ the effective
action in the scaling region
suffers only the anomaly and no further symmetry breaking.
There is no reason, of course, for the numerical values of
$C^{AA}$ and $C^{VA}$ to be the same as in sect.~2, where we used
the spherical cut-off as a regulator.

The above counterterms may be extended to periodic functions in
$A_{\mu}$ and $V_{\mu}$. E.~g. writing $C^{AA}_{\mu\nu} = \ta
\dl_{\mu\nu}$ the $C^{AA}$ term may be replaced by the standard
lattice form for a gauge field mass term $\ta \sum_{x\mu} (1-\cos
A_{\mu x})$. This replacement could help to reduce the scaling violations.
In this paper we shall however stay with the form (\ref{CT}).

The currents $J^{V,A}_{\mu}$ are identified from the total action
$S+S_{ct}$ by letting $A_{\mu} \ra A_{\mu} +\dl A_{\mu}$,
$V_{\mu} \ra V_{\mu} + \dl V_{\mu}$, and collecting terms linear
in $\dl A_{\mu}$ and $\dl V_{\mu}$. The current correlation
functions are obtained by differentiating the effective action
with respect to $A_{\mu}$ and $V_{\mu}$. As we shall study
$\langle J^{V,A}_{\mu}\rangle$ only for zero $V_{\mu}$
(but for arbitrary $A_{\mu}$), we give the currents here for
$V_{\mu}=0$,
\bea
 J^V_{\mu x}  \!\!\!&=&\!\!\! \frac{i}{2} \Tr \left[
\Psb_x \g_{\mu} (U_{\m x}P_L+U^*_{\m x} P_R) \Psi_{x+\mh} +
\Psb_{x+\hmu} \g_{\m} (U_{\m x}^* P_L + U_{\m x} P_R) \Psi_{x}
\right] +
C^{VA}_{\mu\nu}A_{\nu x}\;, \label{VCURR}
\\
 J^A_{\mu x}  \!\!\!&=&\!\!\!  \frac{-i}{2} \Tr \left[
\Psb_{x} \g_{\m} (U_{\mu x} P_L-U^*_{\m x}P_R) \Psi_{x+\mh} +
\Psb_{x+\hmu} \g_{\m} (U^{*}_{\mu x} P_L-U_{\m x} P_R)
\Psi_{x} \right] +
C^{AA}_{\mu\nu} A_{\nu x}\; ,  \nonumber \\
& & \label{ACURR}
\eea
where $U_{\m x} = U_{\mu x}^L = \exp(-iA_{\mu x})$.
A natural choice for the pseudoscalar density is given by
\be
 J^P_{x}= i\Tr \left[\Psb_x \g_5 \Psi_x \right] \;. \label{PCURR}
\ee

The currents in terms of the
staggered field $\chi$ are obtained by
inserting the relations (\eq{CHIPSI})
into (\eq{VCURR})-(\eq{PCURR}),
\bea
J_{\mu x}^V \!\!&=&\!\! \frac{i}{8} \left[ c_{\mu x} \sum_b
\et_{\m x+b} (\cb_{x+b} \c_{x+b+\hm} + \cb_{x+b+\hm}
\c_{x+b}) \right. \nonumber\\
&& \left. -s_{\mu x} \sum_{b+c=n}
\et_{12 x+c}(\et_{\mu x+c} \cb_{x+b}\c_{x+c+\hm}
+ \et_{\mu x+b}\cb_{x+b+\hm}\c_{x+c}) \right]
+ C^{VA}_{\mu\nu}A_{\nu x} \;,\lb{LJV}
\\
 J_{\mu x}^A \!\!&=&\!\! \frac{1}{8} \left[ c_{\mu x} \sum_{b+c=n}
\et_{12 x+c} (\et_{\mu x+c}\cb_{x+b}\c_{x+c+\hm} -
              \et_{\mu x+b}\cb_{x+b+\hm}\c_{x+c}) \right. \nonumber\\
&& \left. - s_{\mu x} \sum_b \et_{\mu x+b}
(\cb_{x+b}\c_{x+b+\hm} - \cb_{x+b+\hm}\c_{x+b}) \right]
+ C^{AA}_{\mu\nu} A_{\nu x}\;, \lb{LJA}
\\
 J_{x}^P \!\!&=&\!\! \frac{1}{4} \sum_{b+c=n}
 \et_{12 x+b}\cb_{x+b}\c_{x+c} \;.\lb{LJP}
\eea
The above vector current $J_{\mu x}^V$ is not directly related to the exact
global U(1) invariance (\eq{EU1}) which we mentioned in the
previous section, because the prescription (\ref{MODAV}) does
not make this symmetry an exact local symmetry.

To obtain the  divergence relation of the conserved current $j_{\mu x}^V$
which is associated with the  exact U(1) symmetry (\eq{EU1})
we replace $\c_x \ra \exp(i\o_x)\c_x$ and $\cb_x \ra \cb_x \exp(-i\o_x)$
in the action (\ref{MODCHI}) and collect terms linear
in $\o_x$. But then the three link couplings in
(\ref{MODCHI}) lead to an awkward looking divergence equation.
The usual form of the divergence equation
is obtained in the canonical model (\ref{CAN}). We find,
\bea
&& \del^{\prime}_{\m} j_{\m}^V=0\;, \nonumber \\
&& j_{\mu x}^V= i \sum_{\mu} \left[ \bar{c}_{\mu x} \et_{\mu x} \half
(\cb_x \c_{x+\hm} + \cb_{x+\hm}\c_x)
-\bar{s}_{\nu x} \ep_{\mu\nu} \et_{\mu x} \half
(\cb_x \c_{x+\hat{\mu}} - \cb_{x+\hat{\mu}}\c_x) \right] \;.\lb{JVE}
\eea

The potential problem of the appearance of additional
global symmetries on the lattice has been emphasized in
particular in ref.~\cc{Ba91}, in the context of fermion number
non-conservation in the Standard Model. We shall assume here the
following resolution, which is, as we believe, in accordance with
current lore \cc{Man}. We can construct many currents, each of which
reduces in the scaling region to a linear combination of the gauge
invariant $J^V_{\mu}$ and the gauge variant $\ep_{\mu\nu}A_{\nu}$.
In particular, the exactly conserved but gauge non-invariant
current $j_{\mu x}^V$ will reduce to the
divergence free combination $J^V_{\mu} -
(i/\pi) \ep_{\mu\nu}A_{\nu}$. Because this current is not gauge
invariant, the corresponding conserved charge is unphysical. It may
have a physical (gauge invariant) component, but there is no reason
why this should be conserved
(for an exposition of the
physics of the equivalent Schwinger model, see for example
ref.~\cc{Ma85}).

In this way the non-gauge invariance of our lattice model
provides presumably a possible way out of the embarrassing
exact global U(1) invariance.
Another possibility to avoid difficulties with
undesired global invariances
is to construct the lattice models such that additional
extra symmetries do not emerge \cc{Sm92,Ba91,EiPr}.
The potential problem of having a larger global
symmetry group on the lattice than in the continuum target model,
requires further detailed investigation.
However, in this paper we shall restrict
ourselves to a study of the currents $J_{\mu x}^V$ and $J_{\mu x}^A$
as given in the eqs.~(\eq{LJV}) and (\eq{LJA}).

With the above definitions of the currents and using the
staggered fermion formalism outlined in ref.~\cc{GoSm,DoSm}
we derived the lattice analogues of the Ward identities in
sect.~2. Let us first concentrate on the vector current.
The diagram in fig.~\ref{FIG1}a gives a non-zero contribution,
whereas the contribution from the typical lattice tadpole
diagram in fig.~\ref{FIG1}b happens to vanish.
The amplitude then reads
\bea
T^{VA}_{\m \n}(p) &=& i e^{i(p_{\n}-p_{\m})/2}
\int_q \frac{ m^2 \e_{\m \n} +\e_{\m \a} s(q_{\a})
  s(q_{\n}+p_{\n})+\e_{\n \a} s(q_{\a}+p_{\a}) s(q_{\m})}
  {D(q) D(q+p)} \nonumber \\
&\times& c(q_{\m}+p_{\m}/2) c(q_{\n}+p_{\n}/2) \prod_j c(p_j/2) c(q_j+p_j/2)\;,
\lb{LTVA}
\eea
where $s(q_j)= \sin(q_j)$, $c(q_j)= \cos(q_j)$, $D(q)=\sum_{\al}
\sin^2 q_{\a} + m^2$
and $\int_q=\int_{-\pi/2}^{+\pi/2} d^2 q/\pi^2$.
To calculate the continuum limit of (\eq{LTVA}) we
let $p$ and $m$ approach zero, and separate the integration region into a
small ball around the origin, $|q|< \dl$, with radius $\dl \ll \pi/2$
and the outer region, $|q| \geq \dl$ (see, for example,
\cc{GoSm}). We let $\dl\ra 0$, $m/\dl\ra 0$ and $p/\dl\ra 0$,
with $p/m$ fixed.
In the inner region we can replace the integrand by its covariant
form, while in the outer region the integrand is expanded in powers
of $m$, $p$ and only the non-vanishing terms are kept.
For the integral (\eq{LTVA}) the contribution from the outer region
vanishes and only the contribution from the inner region remains,
which is exactly a continuum loop integral with a spherical cut-off
$\dl$, and coincides with eq.~(\eq{TVA1}), except for a factor of
two corresponding to the two flavors.
Consequently, after normalizing to one flavor we have the same contact
term (\ref{C1}) as in the cut-off regulated continuum theory.
So we find the following vector current divergence relation on the
lattice
\be
\del^{\prime}_{\mu} \av{ J^V_{\mu x} }_{\ch} =  2i F_x/2\pi
+ O(a) \;, \label{LV}
\ee
where the contact term $C_{\mu \nu}^{VA}$ in the definition (\eq{LJV})
is given in eq.~(\eq{C1}) and $F_x$ is a suitable form for the
field strength, c.f. eq.~(\eq{FHYC}) below.
Here $\del^{\prime}_{\mu} f_{\mu x} \equiv \sum_{\m}
(f_{\mu x}-f_{x-\hm})$ is the divergence on the lattice and
$\av{\bullet}_{\ch}$ denotes the integration over the $\c$ fields
and includes here and in the following also a normalization to one
staggered flavor. The $O(a)$ indicates
terms which arise due to the discretization and vanish when
$a\ra 0$.

For the amplitude $T^{AA}_{\m \n}$ we obtained the expression,
\bea
T^{AA}_{\m \n} (p) \!\!\! &=& \!\!\! T^{AA (a)}_{\m \n}(p)+T^{AA (b)}_{\m
\n}(p)
+ C^{AA}_{\mu\nu} \;, \lb{LTAA} \\
T^{AA (a)}_{\m \n} (p) \!\!\! &=& \!\!\! - e^{i(p_{\n}-p_{\m})/2}
\int_q \frac{ m^2 \dl_{\m \n}+s(q_{\m})
  s(q_{\n}-p_{\nu})-\e_{\m \a} \ep_{\n \b} s(q_{\a}) s(q_{\b}+p_{\b})}
  {D(q) D(q+p)}  \nonumber \\
\!\!\!&\times&\!\!\!
c(q_{\m}+p_{\m}/2) c(q_{\n}+p_{\n}/2)\prod_j c(q_j+p_j/2)^2 \;, \lb{LTAAA} \\
T^{AA (b)}_{\m \n} (p) \!\!\! &=& \!\!\! -\dl_{\m \nu} \int_q \frac{
  s(q_{\m})^2 }{D(q)} \;,\lb{LTAAB}
\eea
where $T^{AA (a)}_{\m \n}(p)$ and $T^{AA (b)}_{\m \n}(p)$ denote the
contributions from the diagrams in fig.~\ref{FIG1}a and b. The
amplitude of the tadpole
diagram is independent of the external momentum $p$.
After a normalization to one flavor, we find
the lattice divergence relation for the axial-vector
current,
\be
     \del^{\prime}_{\mu} \av{J^A_{\mu x} }_{\ch}
     = 2m \av{ J^P_{x} }_{\ch}   + O(a) \;,
                                    \label{LA} \\
\ee
where the contact term coefficient $C^{AA}_{\mu \nu}$
in (\eq{LJA}) has to be chosen as
\bea
&& C^{AA}_{\mu \nu}=\tau \dl_{\m \n} \;, \;\;
    \tau=-\left( \frac{1}{2\pi}+I_1+I_2 \right) \approx 0.0625 \;, \lb{C2L} \\
&& I_1= \frac{1}{2} \int_q \frac{[s(q_2)^2 -s(q_1)^2] c(q_1)^4
  c(q_2)^2 }{\left[s(q_1)^2 + s(q_2)^2 \right]^2 }
  \approx 0.0283 \;, \;\;
 I_2=- \frac{1}{2} \int_q \frac{ s(q_1)^2 }
{\left[s(q_1)^2 + s(q_2)^2 \right]^2 }
  =-\frac{1}{4}
\;. \nonumber \\
&& \lb{INT}
\eea
The first two terms, $1/2\pi$ and $I_1$,
in (\eq{C2L}) come from inner and outer
region parts of the integral (\eq{LTAAA}) and $I_2$ from the
lattice integral (\eq{LTAAB}).

An alternative form of the staggered fermion action is given by
eq.~(\eq{MODCHI}), but with non-compact gauge fields, i.~e. with
the replacements $c_{\m x} \ra 1$ and
$s_{\m x} \ra A_{\m x}$. Of course, also this form reduces in classical
continuum limit to the target model with two flavors.
Using this modified form of the action
we would have to drop the $U$ fields from eq.~(\eq{ACURR}).
Then the tadpole diagram in fig.~\ref{FIG1}b would not have
given a contribution and we would instead have obtained
$C_{\m \n}^{AA}\approx -0.1875 \; \dl_{\m \nu}$ which is
larger than the result in (\eq{C2L}).
This shows that the form (\eq{MODCHI})
is more appropriate for restoring gauge invariance.
However, unlike with the continuum point
split current \cc{GrJa72}, a contact term is still
needed because the staggered theory is not gauge invariant.
\section{Numerical results}
In this section we numerically compute the current divergence relations
(\eq{LV}) and (\eq{LA}) for smooth external gauge field configurations
with variable amplitude. A similar test \cc{JA} has been recently applied
to a proposal using domain wall fermions \cc{KA}.
Since gauge invariance is violated in the staggered model it is
interesting to investigate the effect of fluctuating gauge
transformations on the divergence relations.
As in an earlier work \cc{SmVi} we shall also investigate the effects
of external gauge fields with $Q \neq 1$.
\subsection{Smooth external gauge field configurations}
To test the relation (\eq{LV}) for the vector current, we use the same
external fields as in ref.~\cc{JA} which are spatially constant,
\be
A_{1 x} = A_0 \sin(2\pi t/T)\;,\;\; A_{2 x} = 0 \lb{CONF1}
\ee
with $t \equiv x_2$. We shall use here and in the following a
lattice, with extents $T$ and $L$ in
the time and spatial directions. The gauge fields $U_{\m x}$ are
close to one everywhere provided the amplitude $A_0$ is
sufficiently small. The topological charge $Q$ is equal to zero
for this class of configurations.

To reduce the discretization error, which arises naturally
when transcribing a continuum gauge field configuration to the
lattice, we use for $F_x$ in (\ref{LV}) the average over
the four lattice plaquettes adjoined to the point $x$:
\be
 F_{x} = \frac{1}{4} \sum_b  F_{12 x-b} \lb{FHYC}
\ee
with $F_{12 x} = \del_1 A_{2x} - \del_2 A_{1x}$ the plaquette field
strength ($\del_{\mu} f_x \equiv f_{x+\hm} - f_x$).
%
%
\begin{figure}
\vspace*{20cm}
\caption{ \noindent {\em
a) The divergence
$-i\del^{\prime}_{\mu} \av{J^V_{\mu x}}_{\ch}$
as a function of $t$ for several values of $A_0$ and $m=0.01$.
The anomaly $F_x/\pi$ is given by the full lines.
b) $\del^{\prime}_{\mu} \av{J^A_{\mu x}}_{\ch}$
as a function of $t$ for several values of $A_0$ and $m=0.01$.
The numerical results for $2 m \av{J_x^P}_{\ch}$
are represented by the full lines which were obtained by connecting
the data points at various $t$ by cubic splines.
}}
\label{FIG2}
\end{figure}
Fig.~\ref{FIG2}a shows the divergence
$-i \del^{\prime}_{\mu} \av{J^V_{\mu x} }_{\ch}$
as a function of $t$ for various values of $A_0$. The quantity
$F_x/\pi$ is represented in this plot by the full lines
which were obtained by connecting the points at $t=1,\ldots,T$ by
a cubic spline. We
use a lattice with $T=L=64$ and periodic boundary conditions
for the $\c$ fields. To regulate the near
zero mode we used a small non-zero mass $m=0.01$. In fig.~\ref{FIG2}a and
in the following graphs we have multiplied the currents and the
anomaly by the lattice volume $L \times T$.
For the smaller amplitudes the agreement between the anomaly and
the current divergence is almost perfect, showing that the
$O(a)$ effects in eq.~(\eq{LV}) are very small.
We find the maximal relative error to increase from $0.1\%$ to
$5\%$ when $A_0$ is raised from $0.1$ to $0.25$.

To test the divergence relation (\eq{LA}) for the axial-vector
current we use the field
\be
A_{1 x} = 0\;,\;\;\; A_{2 x}= A_0 \sin( 2\pi t/T )\;,\lb{CONF2}
\ee
which in contrast to the previous case is longitudinal,
i.e. $\ep_{\m \n} \del_{\mu} A_{\nu} =0$ and
 $\del^{\prime}_{\mu} A_{\mu} \neq 0$.
In fig.~\ref{FIG2}b, $\del^{\prime}_{\mu} \av{J^A_{\mu x} }_{\ch}$
is represented by the various symbols, which as in fig.~\ref{FIG2}a
correspond to different values of the amplitude $A_0$. The quantity
$2 m \av{J_x^P}_{\ch}$ is represented by the solid lines which were
obtained by connecting the data points at $t=1,\ldots,T$ by cubic splines.
The plot shows that the relation (\eq{LA}) holds nicely within the
given range of $A_0$ values. The relative error is smaller than $6\%$.
The complicated looking $A_0$-dependence of
$\del^{\prime}_{\mu} \av{J^A_{\mu x} }_{\ch}$
shows that for $m \ne 0$ the effective action $S_{eff}(A)$,
which results after carrying out the $\c$ integration in the
path integral, and thereto also $\lag J_{\mu x}^A \rag=-\dl
S_{eff}(A)/\dl A_{\mu x}$, are non-linear functionals
of the vector potential $A_{\mu x}$.
\subsection{Fluctuating gauge degrees of freedom}
Since the model lacks gauge invariance it is instructive to
investigate how relation (\eq{LV}) changes if we
perform gauge transformations
\be
U_{\mu x} \ra \Om_x U_{\m x} \Om_{x+\hm}^* \;, \label{OMU}
\ee
on a smooth link configuration.
To investigate the effect of $\Om$ field fluctuations, we can
again compute the current divergence, but now averaged over
an $\Om$ field ensemble,
\be \av{ \del^{\prime}_{\mu} J^V_{\mu x}}_{\Om} \equiv \frac{1}{Z}
\int D\Om\, \del^{\prime}_{\mu} \av{J^V_{\mu x}}_{\ch}
e^{S(\Om)}\;,\;\;\;\; Z=\int D\Om e^{S(\Om)} \;,\label{DELV}
\ee
with Boltzmann weight $\exp S(\Om)$.
We will generate here the configurations of the gauge degrees of freedom
in two different ways, depending on which formulation of the full
model with dynamical gauge fields we are aiming for.

As we mentioned in sect.~1,
one possibility is to use non-perturbative gauge fixing and add
counterterms to restore gauge invariance \cc{Rome89}.
With non-compact U(1) gauge fields (but still coupled
compactly to the fermions), we may use a gauge fixing action
$-\zt /2 \sum_x \{ \del^{\prime}_{\mu} A_{\mu x} \}^2$. The corresponding
Fadeev-Popov determinant is independent of $A_{\mu}$. Since
$A_{\mu x}$ transforms into $A_{\mu x} + \del_{\mu} \om_x$ under
a gauge transformation, this suggests to use the gauge fixing action
\be S(\Om) = -\zeta/2 \sum_x \left\{ \Box \om_x
\right\}^2 \;, \;\;\;\; \Om_x=\exp i
\om_x \;, \label{OM1}
\ee
to generate the gauge degrees of freedom. Here $\Box$ denotes the
lattice laplacian operator $\del^{\prime}_{\mu}\del_{\mu}$ and
$\zeta$ is the gauge fixing parameter ($\zeta=\infty$ corresponds
to the Landau gauge). The non-compact
phases $\om_x$ are coupled through $\Om_x=\exp i \om_x$
and the replacement (\ref{OMU}) to the fermions.
In fig.~\ref{FIG3} the crosses represent the results
for  $-i \av{\del^{\prime}_{\m}J^V_{\m x}}_{\Om}$ after
averaging over 1400 independent $\Om$ configurations at $\zeta=10$.
The numerical result lies slightly below the solid line
which represents here again the anomaly $F_x/\pi$
for the given external gauge field configuration (\eq{CONF1}).
This shows that the anomaly relation
remains valid after multiplying the vector
current by a factor which is slightly larger than one.
Such a current renormalization is expected
because there is no protection by symmetry.
This result shows that a further numerical investigation of the gauge
fixing approach may be technically feasible, at least for U(1) with
non-compact gauge potentials where one
has not to worry about the Fadeev-Popov factor.

We find that at smaller values of $\zeta$ the statistical
fluctuations increase tremendously. Recall, however, that in
the convention we are using the gauge coupling $g$ is absorbed in
$A_{\mu x}$, implying that $\zt\propto 1/g^2$; e.g. $\zt=1/g^2$
corresponds to the Feynman gauge. Since $g^2\ra 0$ in lattice
units, large $\zt$'s are not unnatural.

An alternative approach to regain a gauge invariant  quantum
model aims at a dynamical restoration of gauge
invariance, by integration over all gauge transformations
\cc{Sm88,SmROME}. In this approach the expected mass counterterm
for the gauge field $2\kp \sum_{x \mu}\RE U_{\mu x}$, which
transforms into $2\kp\sum_{x\mu} \RE (\Om_x U_{\mu x}\Om_{x+\hm}^*)$,
suggests to use the action
\be
S(\Om) =  \k \sum_{x\m } \left\{
\Om_x^* \Om_{x+\hmu}+ \Om_{x+\hmu}^* \Om_{x} \right\} \;,
\label{OM2}
\ee
to generate the $\Om$ configurations in (\eq{DELV}). This action is
identical to the action for the XY-model in two dimensions with
hopping parameter $\k$. The most ambitious scenario \cc{Sm88,SmROME}
corresponds to choosing $\kp$ in the vortex phase of the XY-model (in
ref.~\cc{SmROME} denoted as `scenario C'). In other scenarios the
models would have to allow an interpretation of $\Om_x$ as a (radially
frozen) Higgs field, for which $\kp$ would have to be close to the
Kosterlitz-Thouless phase transition at $\kp_c\approx 0.5$.

For large values of $\kp$, deep in the spin wave phase, the fluctuations
of $\Om$ are small and we expect that
$-i\av{\del^{\prime}_{\mu} J^V_{\m x}}_\Om = F_x/\pi$
within statistical errors.
This is confirmed by a simulation at $\kp=2$. For smaller values of
$\kp$, when approaching the phase transition at $\kp_c\approx 0.5$, the
fluctuations increase dramatically. The result at $\kp=1.0$ (indicated
by circles in fig.~\ref{FIG3}) shows that the values for
$-i\av{ \del^{\prime}_{\m} J^V_{\m x}}_{\Om}$
now lie significantly below
$F_x/\pi$ (solid line).
This result was obtained after averaging over 6400 independent
$\Om$ configurations. The dashed line was obtained by fitting the numerical
data to the ansatz
$-c (A_0 2\pi/T)\cos (2\pi t/T)$ with free parameter $c$. The
good quality of the fit shows that the relation (\eq{LV}) remains
valid after renormalizing the vector current
by a factor $1/c \approx 1.20$.
In the vortex phase the fluctuations of
$-i  \del^{\prime}_{\m} J^V_{\m x}$
were so strong that even after an excessive increase of the statistics
we were not able to get a meaningful estimate of the average current
divergence.
The danger is that the nice scaling behavior of the staggered
fermions, which we could demonstrate for smooth external fields,
is washed out when the gauge degrees of freedom
fluctuate too strongly.

An interesting difference between the actions (\ref{OM1})
and (\ref{OM2}) is that in the latter case the small fluctuations
in the spin wave phase are weighted by $\exp(\kp \om\Box\om)$,
whereas in the former case a $\Box^2$ appears instead of the $\Box$,
which leads to much smoother $\om$ configurations.
%
%
\begin{figure}
\vspace*{10cm}
\caption{ \noindent {\em
The divergence  $-i \av{\del^{\prime}_{\m} J^V_{\m x}}_{\Om}$ as a
function of $t$ with the gauge degrees of freedom $\Om$ generated by
the gauge fixing action ({\protect \eq{OM1}})
at $\zeta=10$ (crosses) and by the XY-model
action ({\protect \eq{OM2}})
at $\k=1.0$ (circles).  The anomaly $F_x/\pi$ is represented by
the full line. The dashed curve is obtained by a fit.
}}
\label{FIG3}
\end{figure}
\subsection{Configurations with $Q \ne 0$}
One would like to promote the divergence equation (\ref{DIV})
to gauge fields with arbitrary topological charge. At first sight
this leads to an apparent contradiction: On a torus with periodic
boundary conditions for gauge invariant quantities, integration
over the left-hand side of eq.~(\eq{DIV}) gives zero, while the
right-hand side gives $2 i Q\neq 0$. This paradox
makes it interesting to see if the
{\em local} divergence equation (\ref{DIV}) is
valid also for topologically non-trivial gauge fields.

In the massive Schwinger model the divergence equation for
the axial-vector
current reads, $\del^{\prime}_{\mu} J^A_{\mu x}=2 m J^P_x+2iq$. After
integration over the torus with periodic boundary conditions
we find $Q=m \Tr ( \gm_5 [\Dsl+m]^{-1})$
and the Atiyah-Singer index theorem is effectively valid \cc{SmVi}.
A paradox is therefore avoided by the fermion mass term in the
massive Schwinger model.
The paradox in the axial-vector model would not emerge
if we would replace the usual mass term in eq.~(\eq{CONT}) by
a Majorana mass term which makes the model equivalent
to the massive Schwinger model.

However, it is not as easy to deal with $Q \neq 0$ in
the axial-vector model as previously in the
Schwinger model (see ref.~\cc{SmVi}). The reason is that we recover
gauge invariance only if the gauge potentials are smooth (and, of course,
$m=0$). For
example, the perturbative derivation of the divergence relations
is valid only when the momenta of the external gauge field
$A_{\mu}$ are negligible compared to the cut-off. On a periodic
lattice a typical lattice gauge field with non-zero topological
charge is not smooth. As an example we investigate the effect of
configurations with constant $F_{12 x}$ used in the vector case
in ref.~\cc{SmVi}.

We consider a configuration with a constant field strength
$F\equiv F_{12 x} = 2 \pi Q/TL$,
\be
\begin{array}{rcll}
U_{1 x} \!\!\!& = &\!\!\! \exp(i F t)     \;,\;\;\;\; & t=1,\ldots,T \;, \\
U_{2 x} \!\!\!& = &\!\!\! 1               \;,\;\;\;\; & t=1,\ldots,T-1 \;,\\
U_{2 x} \!\!\!& = &\!\!\! \exp(i F T x_1) \;,\;\;\;\; & t=T \;.
\end{array}
\lb{CONF3}
\ee
For small $Q$ the link field $U_{\m x}$ is close to one and smooth
everywhere except
for $t=T$ where $U_2$ contains a transition function \cc{SmVi}.
When shifting $t$ from $T-1$ to $T$ and then to $T+1=1$ (mod $T$)
$U_2$ makes a jump as a function of $t$ (e.g. for
$Q=1$ and $x_1=L/2$ $U_2$ jumps from $+1$ to $-1$ and then to $+1$ again).
Also $U_1$ jumps when shifting $t$ from $T$ to $t=T+1=1$ (mod $T$).
In gauge invariant models these transition functions
are invisible.

We expect the divergence equation to hold with small $O(a)$
corrections in the region of the lattice where the gauge fields
are sufficiently smooth, but expect deviations in the region
near the transition function.
As in fig.~\ref{FIG2} we have plotted in fig.~\ref{FIG4}
the divergence $-i  \del^{\prime}_{\mu} \av{J^V_{\mu x}}_{\ch}$
(squares) for the
space slice $x_1=L/2$ as a function of $t/T$ with $L=T=64$ (diamonds)
and $L=T=32$ (crosses), but now for the
gauge field configuration (\eq{CONF3}). We used here
antiperiodic boundary conditions for the $\c$ fields and $m=0$.
For other space slices we got similar plots.
The solid line represents again $F_{x}/\pi$,
which now is independent of $x$. Far away from the time slice
which carries the transition function the agreement of
$-i  \del^{\prime}_{\mu}  \av{J^V_{\mu x}}_{\ch}$ with $F_{x}/\pi$
is satisfactory, however in the vicinity of this time slice
the deviations become huge (the values
at $t=31,32$ ($T=32$) and $t=62,63,64$ ($T=64$) were
dropped from the graph since they are much larger than $20$).
The figure indicates also that the region of disturbance shrinks for
increasing lattice size.
The strong deviations induced by the transition function
is analogous to the enormous fluctuations
we observed in $-i  \del^{\prime}_{\mu} \av{J^V_{\mu x}}_{\Om}$
when lowering $\k$ in the action (\ref{OM2}).
The increase of the vorticity renders the effective
gauge field configuration $\Om_x U_{\m x} \Om_{x+\hmu}^*$ very
rough, similar to the above gauge field at $t=T$.

%
%
\begin{figure}
\vspace*{10cm}
\caption{ \noindent {\em
The divergence $-i  \del^{\prime}_{\mu} \av{J^V_{\mu x}}_{\ch}$
as a function of $t/T$ for a configuration with $Q=1$ on two different
lattices ($T=L=32$, $64$). We have used $m=0$,
and antiperiodic boundary conditions for the $\c$ field.
The solid line represents the anomaly $F_x/\pi$.
}}
\label{FIG4}
\end{figure}
The above results for the $Q=1$ configuration are not entirely
satisfactory.
To avoid the large errors with topological non-trivial gauge
fields we will have to follow the mathematicians and introduce
charts in which the gauge potentials are smooth,
such that the Dirac operator has negligible discretization errors
in any open region (see e.~g. ref.~\cc{Na}).
A proper treatment of this will have to be done
in the future.
\section{Discussion}
We have showed that the staggered fermion model (\eq{MODCHI})
can reproduce continuum  Ward identities
after incorporating the appropriate counterterms.
The coefficients of these counterterms have been computed in this paper
within lattice perturbation theory.
Using smooth external gauge fields with zero topological charge
we have numerically verified the validity of the
divergence relations for the vector and axial-vector currents
and found good accuracy,
for gauge field configurations with $|U_{\m x}-1| \aleq 0.2$.

We tested the sensitivity of the divergence relation for the
vector current to fluctuating gauge degrees of freedom
which were generated either with
the gauge fixing action (\eq{OM1}) or the scalar field action (\eq{OM2}).
After renormalizing the currents with a finite factor $>1$,
the divergence relation remains valid in both cases, provided that the
fluctuations of the gauge modes $\Om_x$ are not too strong.
This is an encouraging result for the gauge fixing approach to chiral
theories. When the gauge modes becomes less constrained,
the induced fluctuations are very severe and might
wash out the anomaly signal completely.

The disconcerting effect of rough gauge transformations is also seen
when the current divergence is measured for a non-smooth gauge field
with topological charge one, used in earlier tests in QED$_2$.
We found the divergence relation for the vector current to be
strongly violated near the region of the lattice where
the gauge potentials lack smoothness.
This means that in a description which violates gauge
invariance at the cut-off level, topologically non-trivial gauge
fields have presumably to be dealt with
the full apparatus of charts
and transition functions that are smooth in space and time.

The results of this paper show that our staggered fermion approach
passes a first test in reproducing the current divergence relations.
It is not clear yet at this stage whether it is possible to
obtain a valid quantum model after the integration over
all gauge field configurations has been carried out in the path integral.
This question shall be addressed in a separate publication \cc{BoSm93}.
Further clarification is also needed of the question how (if indeed)
the additional global U(1) invariance on the lattice
does not give rise to a local conservation law.\\

\noindent {\bf Acknowledgements:}
The numerical computations were performed on the CRAY Y-MP4/ 464
at SARA, Amsterdam. This research was supported by the ``Stichting voor
Fun\-da\-men\-teel On\-der\-zoek der Materie (FOM)''
and by the ``Stichting Nationale Computer Faciliteiten (NCF)'' and by
the DOE under contract DE-FG03-91ER40546.\\
%
%
%

\end{document}